\documentclass[usenatbib]{mn2e}
\usepackage{graphicx}

\def\apj{\rm ApJ}

\def\aj{\rm AJ}
\def\mnras{\rm MNRAS}

\def\pasa{\rm PASA}
\def\aap{\rm AAP}
\def\araa{\rm ARA\&A}

\def\gax{\mathrel{\raise.3ex\hbox{$>$}\mkern-14mu\lower0.6ex\hbox{$\sim$}}}
\def\lax{\mathrel{\raise.3ex\hbox{$<$}\mkern-14mu\lower0.6ex\hbox{$\sim$}}}
\def\gtorder{\mathrel{\raise.3ex\hbox{$>$}\mkern-14mu
             \lower0.6ex\hbox{$\sim$}}}
\def\ltorder{\mathrel{\raise.3ex\hbox{$<$}\mkern-14mu
             \lower0.6ex\hbox{$\sim$}}}

\begin{document}

\title [On The Red Supergiant Problem]
   {On The Red Supergiant Problem}

\author[C.~S. Kochanek]{ 
    C.~S. Kochanek$^{1,2}$ 
    \\
  $^{1}$ Department of Astronomy, The Ohio State University, 140 West 18th Avenue, Columbus OH 43210 \\
  $^{2}$ Center for Cosmology and AstroParticle Physics, The Ohio State University,
    191 W. Woodruff Avenue, Columbus OH 43210 \\
   }

\maketitle

\begin{abstract}
We examine the problem of estimating the mass range 
corresponding to the observed red supergiant (RSG) progenitors of Type~IIP
supernovae.  Using Monte Carlo simulations designed to reproduce the 
properties of the observations, we find that the approach of 
\cite{Davies2018} significantly overestimates the maximum mass, 
yielding an upper limit of $M_h/M_\odot = 20.5 \pm 2.6$ for an 
input population with $M_h/M_\odot = 18$.  Our preferred Bayesian 
approach does better, with $M_h/M_\odot = 18.6 \pm 2.1$ for the same 
input populations, but also tends to overestimate $M_h$.  For the actual
progenitor sample and a Salpeter initial mass function we find
$M_h/M_\odot = 19.01_{-2.04}^{+4.04}$ for the \cite{Eldridge2004}
mass-luminosity relation used by \cite{Smartt2009} and 
\protect\cite{Davies2018}, and $M_h/M_\odot = 21.28_{-2.28}^{+4.52}$
for the \cite{Sukhbold2018} mass-luminosity relation.  Based
on the Monte Carlo simulations, we estimate that these are
overestimated by $(3.3\pm0.8)M_\odot$.  The red supergiant 
problem remains. 
\end{abstract}


\section{Introduction}

Particularly as the archive of Hubble Space Telescope images of nearby 
galaxies has grown, there has been steady progress in identifying the
progenitors of core collapse supernovae (ccSNe, see the reviews by
\cite{Smartt2009} and \cite{Smartt2015}).  In \cite{Kochanek2008},
we pointed out that there appeared to be a deficit of higher mass
progenitor stars.   This point was made more cleanly 
and with better statistics for Type~IIP
ccSNe by \cite{Smarttetal2009}.  The progenitors of Type~IIP ccSNe
are red supergiants, and \cite{Smarttetal2009} found progenitors
with masses between $8.5^{+1.0}_{-1.5} M_\odot$ and 
$(16.5 \pm 1.5)M_\odot$, while red supergiants in the local group
are found with masses of up to $25 M_\odot$.  \cite{Smarttetal2009}
termed this the ``red supergiant problem.'' 

The red supergiant problem could be solved by eliminating
the gap between the highest observed progenitor masses 
and the predicted maximum masses at which
stars explode as red supergiants.  One possibility is to 
modify stellar evolution and mass loss to reduce the maximum
mass of stars exploding as RSGs and have them instead explode
as Type~Ib or Type~Ic ccSNe (\citealt{Groh2013}).
A second possibility is to posit that the luminosities,
and hence the masses, of the observed progenitors have been
underestimated due to unrecognized foreground or circumstellar
extinction (\citealt{Walmswell2012}, \citealt{Beasor2016}, \citealt{Davies2018}). Note,
however, that it is easy to overestimate the effects of dust
(\citealt{Kochanek2012}).  A third possibility is to
argue that the difference between the maximum masses of progenitors
and the expected maximum masses of RSGs is statistically 
insignificant (\citealt{Davies2018}).

The alternative physical explanation is that the missing progenitors are not being
found because the more massive RSGs are not exploding as
SNe and instead become black holes (\citealt{Kochanek2008}).
Stars in the mass range of the missing RSG progenitors have
internal structures that are particularly difficult to 
explode (\citealt{OConnor2011}, \citealt{Ugliano2012},
\citealt{Pejcha2015}, \citealt{Sukhbold2016}) and failed
explosions of these RSGs provide the first natural explanation
for the observed masses of Galactic black holes 
(\citealt{Kochanek2014}, \citealt{Kochanek2015}).
Furthermore, our search for failed SNe with the Large
Binocular Telescope (\citealt{Gerke2015}, \citealt{Adams2017a},
\citealt{Adams2017b}) has identified one excellent candidate 
for a failed ccSN
whose estimated progenitor mass is exactly in the range 
needed to explain the RSG problem and the masses of the
Galactic black holes.  The failed ccSNe rate implied by the discovery
of one candidate is also consistent with theoretical expectations.

In this paper we reconsider the problem of estimating the mass
range of RSG progenitors.  We assume that  
stars explode in a mass range from $M_l$
to $M_h$ with a \cite{Salpeter1955} power law initial
mass function (IMF), $dn/dM \propto M^{-1-x}$ with $x=1.35$.
In modern
examinations of the explodability of stars (e.g., \citealt{OConnor2011}, \citealt{Ugliano2012},
\citealt{Pejcha2015}, \citealt{Sukhbold2016}), the relationship
between mass and outcome is more complex, with explosions and
failures interspersed in mass, but there is still effectively
a maximum mass.  The objective is to estimate the two
mass limits $M_l$ and $M_h$.  In particular, \cite{Davies2018}
carries out an analysis to find an upper limit of 
$M_h/M_\odot=19.0_{-1.3}^{+2.5}$ that is significantly above the
estimate of $M_h/M_\odot=16.5\pm1.5$ by \cite{Smartt2009}. 
\cite{Davies2018} further argue that this should be corrected
to $M_h/M_\odot=25$ because the highest mass observed 
progenitor must lie below $M_h$ leading to an underestimate 
of the limit that requires an upward correction. 

A simple way to examine this question is to use Monte Carlo
simulations designed to closely mimic the properties of the
observations and then analyze them to see how well the input
mass limits are recovered.  We will consider both the 
\cite{Davies2018} analysis method and a Bayesian approach
that is similar in spirit to the original \cite{Smartt2009}
analysis.  In \S2 we describe the calculations and in 
\S3 we discuss the results.     

\section{Methods}

For this paper, we simply adopt the tabulation of the properties
of 24 Type~II progenitors from \cite{Davies2018}.
The progenitors are characterized by a distance modulus,
$\mu$, a broad band filter magnitude or magnitude limit, $m_\lambda$,
an estimated extinction for that wavelength, $A_\lambda$,
and a bolometric correction, $BC_\lambda$.  \cite{Davies2018}
treat SN~2009md slightly differently, but we filled in the
missing values in their Table~4 so as to reproduce their
estimates of the progenitor luminosity and its uncertainties.  
Associated with
each quantity is an uncertainly: $\sigma_\mu$, $\sigma_m$,
$\sigma_A$ and $\sigma_{BC}$.  We also require statistical
distributions for these quantities. \cite{Davies2018} treat the
distributions as Gaussians except for the bolometric 
correction, which is viewed as uniformly distributed between
$BC_\lambda-\sigma_{BC}$ and $BC_\lambda+\sigma_{BC}$.   
\cite{Davies2018} round negative extinctions in the tails
of the Gaussian extinction distribution upwards to zero.
There are 14 flux measurements and 10 upper limits.
Where there are flux limits, they are all $3\sigma$ limits,
with the exception of a $5\sigma$ limit for SN~2002hh.
As a slight simplification, we convert this into a $3\sigma$
limit so that all the limits can be treated uniformly.

\begin{figure}
\centering
\includegraphics[width=0.50\textwidth]{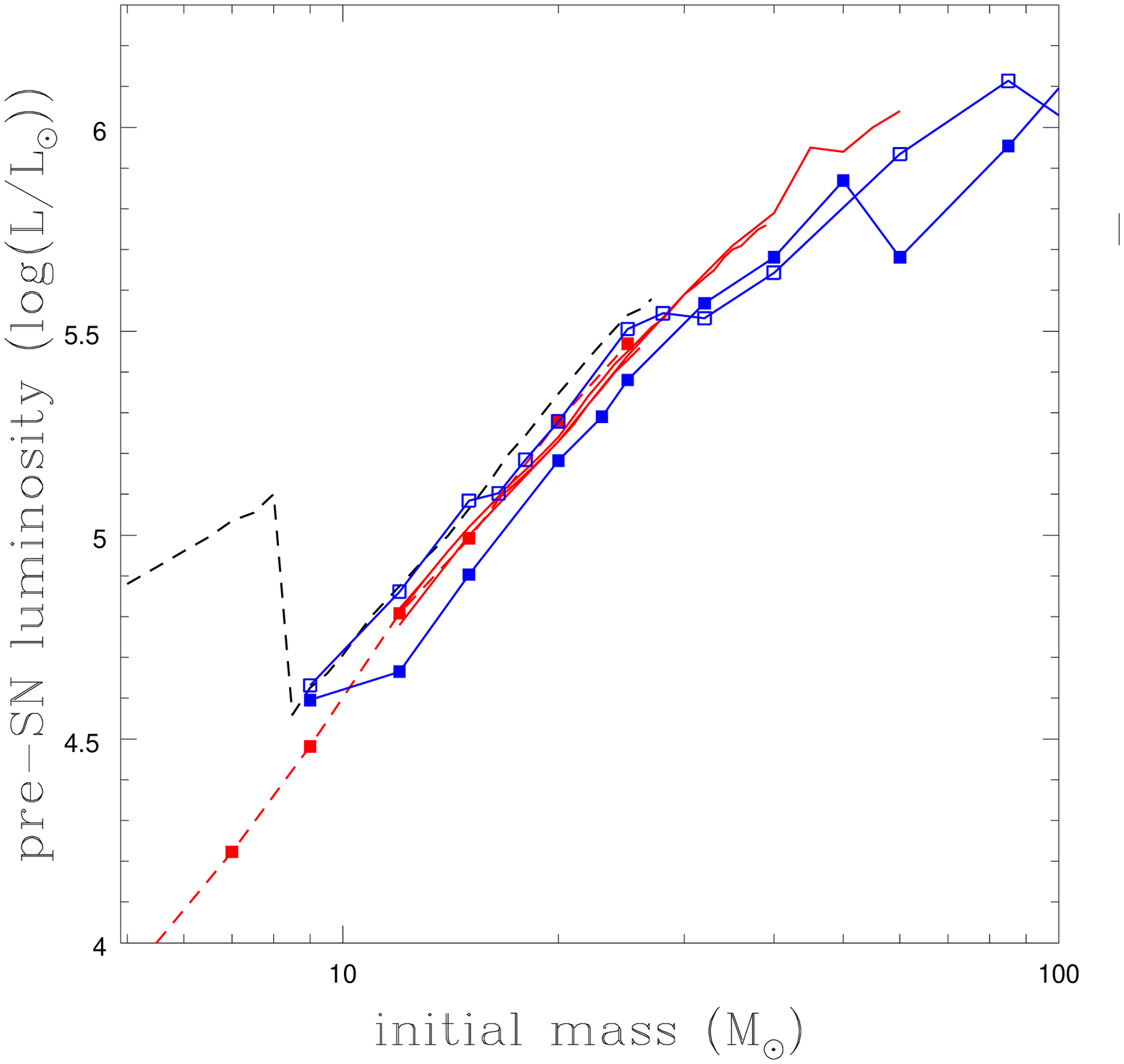}
\caption{
  End of life mass-luminosity relations from \protect\cite{Schaller1992}
  (red dotted, filled squares), \protect\cite{Eldridge2004} (black dashed),
  non-rotating \protect\cite{Groh2013} (blue solid, filled squares), rotating
  \protect\cite{Groh2013} (blue solid, open squares), and \protect\cite{Sukhbold2018}
  (red solid).  The three mass loss models from \protect\cite{Sukhbold2018} 
  lie almost on top of one another.  Only  the \protect\cite{Eldridge2004}
  models include high luminosity AGB phase at lower masses.
  }
\label{fig:mass}
\end{figure}

Given these quantities, the progenitor luminosity $L$ is
\begin{equation}
   2.5 \log \left( {L \over 78.6 L_\odot} \right) = -m_\lambda + \mu + A_\lambda - BC_\lambda.
   \label{eqn:maglum}
\end{equation}
The data really only determines a minimum and maximum progenitor
luminosity, but this can be converted to a mass range given
a mass-luminosity relation.  Figure~\ref{fig:mass} shows the end of 
life mass-luminosity
relations from \cite{Schaller1992}, \cite{Eldridge2004},
\cite{Groh2013} and \cite{Sukhbold2018}.  \cite{Smartt2009}
and \cite{Davies2018} primarily used the \cite{Eldridge2004}
models after eliminating the luminous AGB phase for lower mass stars.
Each model has some mass above which the models cease to be
RSGs at death.  \cite{Sukhbold2018} includes models with 
their standard mass loss rate, half that rate and one tenth
of that rate, with the stars remaining as RSGs up to $26$,
$39$ and $60M_\odot$.  In Figure~\ref{fig:mass}, these three
mass loss sequences are virtually indistinguishable, essentially because
the mass of the envelope has no effect on the luminosity of the helium core.  

For our calculations, we need a mass-luminosity relation that extends
beyond the mass range assumed to explode as an RSG, so we use
the low mass loss models from \cite{Sukhbold2018} extended to
lower mass ($<12M_\odot$) using the models from \cite{Schaller1992}
since the two sets of models overlap.  For ease of calculation,
\begin{equation}
  \log { L \over L_\odot} = 4.610 + 2.267\log\left({ M\over 10M_\odot}\right) 
       -0.494 \log^2\left({M \over 10M_\odot}\right)
      \label{eqn:mgivesl}
\end{equation} 
and 
\begin{equation}
  \log { M \over M_\odot} = 1.180 + 0.489\log\left({L \over10^5L_\odot}\right) 
             +0.056 \log^2\left({L\over 10^5L_\odot}\right)
\end{equation} 
provide very good polynomial fits to the resulting mass-luminosity 
relation for $5M_\odot < M < 60 M_\odot$.  The shape of the polynomials 
also fits the \cite{Eldridge2004} 
models well, but the leading constants become $4.703$ and $1.131$ for the
luminosity and mass, respectively. The offsets mean that the
\cite{Sukhbold2018} models are 24\% less luminous at fixed mass
and 12\% more massive at fixed luminosity than the \cite{Eldridge2004}
models.  

Like \cite{Davies2018} we simply assume a \cite{Salpeter1955}
IMF, $dN/dM \propto M^{-x-1}$ with $x=1.35$ leading to an integral
distribution of progenitor masses of
\begin{equation}
    P_{SN}(<M) = { M_l^{-x} - M^{-x} \over M_l^{-x} - M_h^{-x} }
    \label{eqn:dist}
\end{equation}
over the mass range $M_l \leq M \leq M_h$. This can then
be inverted to get the mass
\begin{equation}
   M_{SN}(P) = M_l \left[ 1 - P\left( 1 - \left( { M_h \over M_l} \right)^{-x}\right)
           \right]^{-1/x},
     \label{eqn:dist2}
\end{equation}
corresponding to fraction
$P$ of the progenitor distribution.  The goal is to
estimate the two mass limits $M_l$ and $M_h$ given the
properties of the progenitors.

We build Monte Carlo test samples
similar to the data as follows.  First, we estimate the $1\sigma$
noise level for each of the 24 SNe.  For the flux limits, this
simply means dividing the stated flux limit by the stated
statistical significance.  For those with measurements, we assume,
as is almost certainly the case, that the noise is background
dominated.  We then convert the progenitor magnitude and its error into
a flux and its error, and the flux error should correspond
to the $1\sigma$ noise level of the data.
Next we assume a minimum and a
maximum mass, where we used $M_l=8M_\odot$ and $M_h=18M_\odot$
or $21M_\odot$,
and then randomly draw a mass using Equation~\ref{eqn:dist2} for
each SN.  This provides a luminosity through Equation~\ref{eqn:mgivesl}, which we convert to an
apparent magnitude by randomly drawing a distance modulus
(Gaussian), extinction (Gaussian rounded up to zero) and
a bolometric correction (uniform) using Equation~\ref{eqn:maglum}.  If the resulting magnitude
is above $3\sigma$, it is treated
as a measurement, and if it is below, we use the 
flux limit instead.  This produces a random sample of
progenitors and limits with the statistical properties
of the data.

\cite{Davies2018} make $10^5$ Monte Carlo trials to estimate
the mass limits.  For each trial, they randomly draw distances
(Gaussian), magnitudes (Gaussian), extinctions (Gaussian rounded up to zero)
and bolometric corrections (uniform) for each SN $i$
to derive a luminosity $L_i$ which is then converted to 
a mass $M_i$. 
For the progenitors with only upper flux limits, the
magnitude is taken to be the stated limit, 
leading to an upper limit on the 
luminosity and mass for the progenitor in the trial.
They then sort the masses and mass limits, discarding
any upper mass limits above the highest mass measurement,
to leave $N'$ objects.  If we index these objects as
$j=0$ to $N'-1$ and define $u_i=1$ for detections and
$u_i=0$ for non-detections, they minimize the statistic
\begin{equation}
    \chi^2 = \sum_{j=0}^{N'} u_j 
   \left[ M_j - M_{SN}\left({ j \over N'-1 } \right) \right]^2,
    \label{eqn:chisq}
\end{equation}
to estimate $M_l$ and $M_h$.  Note that only the detections
($u_j\equiv 1$)
contribute to the statistic, with the highest mass detection 
having $M_{SN}=M_h$.  The lowest mass detection or upper limit
has $M_{SN}=M_l$.  The distribution of the resulting $10^5$ values of the
$M_l$ and $M_h$ that minimize this fit statistic
for each Monte Carlo trial provides their estimate of 
the allowed minimum and maximum progenitor masses.  Since
the masses are just weighted uniformly in the $\chi^2$, and only the 
maximum likelihood estimates of $M_l$ and $M_h$ are used
from each trial, there is no need to define the usual
error term in the denominator of the $\chi^2$.   

We prefer a more Bayesian approach that is similar to the
original procedures of \cite{Smarttetal2009}, although we
will keep the same input data and the relations between
fluxes and luminosities as used by \cite{Davies2018}.
We first construct the relative probability distribution
that progenitor $i$ has mass $M$ given the data
$d$.  For a source with a detection, we compute  
\begin{eqnarray} 
   P_i(M_i|d) &\propto  &\int d\mu dA_\lambda dBC_\lambda
        P(\mu) P(A_\lambda) P(BC_\lambda) \\
        && \exp\left( - { \left( m_{mod} - m_\lambda \right)^2 \over 2 \sigma_m^2 } \right), \nonumber
\end{eqnarray}
where the model magnitude $m_{mod}$ comes from rearranging Equation~\ref{eqn:maglum}.
For the upper limits, we compute the probability given the mass that the flux
would not exceed the $3\sigma$ flux limit, 
\begin{eqnarray} 
   P_i(M_i|d) &\propto & \int d\mu dA_\lambda dBC_\lambda
        P(\mu) P(A_\lambda) P(BC_\lambda) \\
        && \hbox{Erfc} \left( { F_{mod}- 3 \sigma_F \over \sqrt{2}\sigma_F} \right)
       \nonumber
\end{eqnarray}
where the various magnitudes must be converted to fluxes and
$\hbox{Erfc}(x)$ is the complementary error function.  $P(\mu)$,
$P(A_\lambda)$ and $P(BC_\lambda)$ are the same probability 
distributions as were used above.  We do not need the 
normalizations of these probability distributions.

Next we must compute the probability of these mass 
probability distributions given the progenitor mass function.
We again use the same fixed relationships between mass and
luminosity.  For each progenitor we need to marginalize over
the luminosity to get
\begin{equation}
   P_i(M_l,M_h|d) \propto \int_{M_l}^{M_h} dM_i P_i(M_i|d) P(M_i|M_l,M_h)
\end{equation}
where $P(M_i|M_l,M_h) = dP(<M)/dM$ is the probability of having
mass $M_i$ given $M_l$ and $M_h$ derived from Equation~\ref{eqn:dist}.
Note that we are maximizing the probability of the detections
having their observed fluxes, and the probability that the 
non-detections are not detected.  The final probability distribution
for the parameters of the mass function is then
\begin{equation}
   P(M_l,M_h|d) \propto P(M_l) P(M_h) \Pi_i P_i(M_l,M_h|d),
\end{equation}
although in practice we compute $\log P(M_l,M_h|d)$ to
avoid floating point underflow problems.  
We use standard logarithmic priors for the mass limits, with
$P(M_l) \propto M_l^{-1}$ and $P(M_h) \propto M_h^{-1}$.
The final distribution is normalized so that 
$\int P(M_l,M_h|d) dM_l dM_h \equiv 1$, and the
distribution for one mass limit is found by 
projecting out the other 
(i.e., $P(M_l|d) = \int dM_h P(M_l,M_h|d)$).

\begin{figure}
\centering
\includegraphics[width=0.50\textwidth]{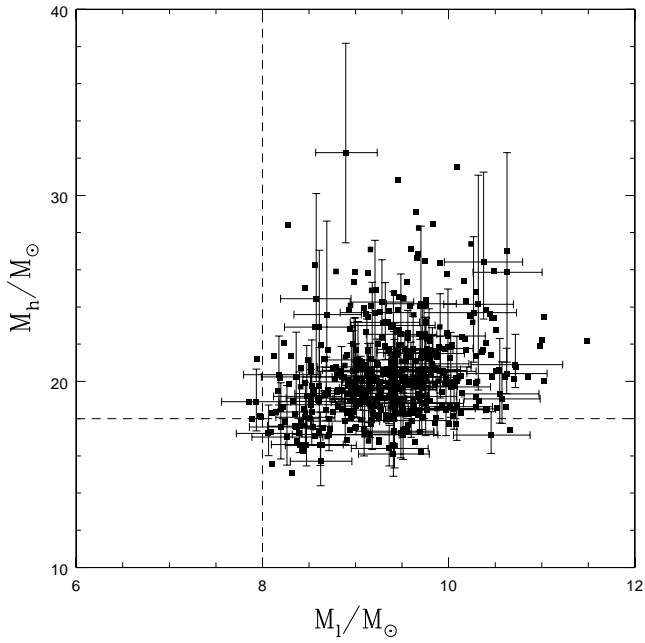}
\caption{
  The results for 500 simulated progenitor data sets using the
  \protect\cite{Davies2018} approach to estimating the
  minimum and maximum progenitor mass.  Each case has
  a point at the median and error bars encompassing 
  68\% (``$1\sigma$'') of the probability are shown
  for 20\% of the trials.  The input values
  are indicated by the dashed lines. 
  }
\label{fig:bplot}
\end{figure}

\begin{figure}
\centering
\includegraphics[width=0.50\textwidth]{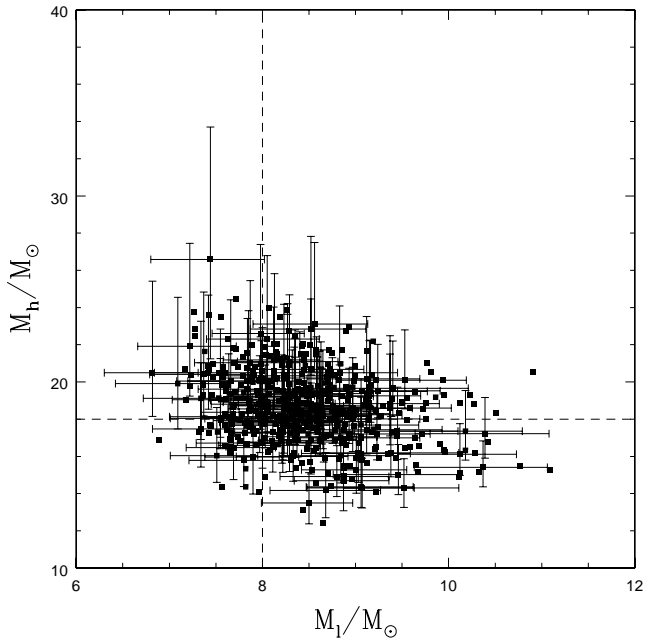}
\caption{
  The results for the same 500 simulated progenitor data
  sets using the Bayesian method presented in \S2. 
  }
\label{fig:cplot} \end{figure}

\begin{figure}
\centering
\includegraphics[width=0.50\textwidth]{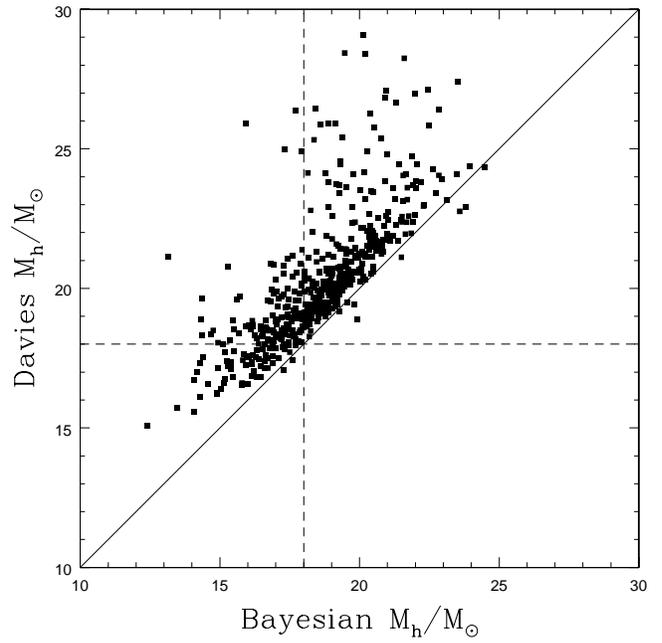}
\caption{
  Estimates of $M_h$ for the Bayesian and
  \protect\cite{Davies2018} methods for each of the
  500 simulated progenitor data sets.  The estimates
  are strongly correlated, but the Bayesian estimates
  are systematically lower and closer to the input
  value.  The dashed
  lines mark the input value of $M_h=18M_\odot$ and
  the diagonal line corresponds to equal mass estimates. 
  }
\label{fig:compare}
\end{figure}

\begin{figure}
\centering
\includegraphics[width=0.50\textwidth]{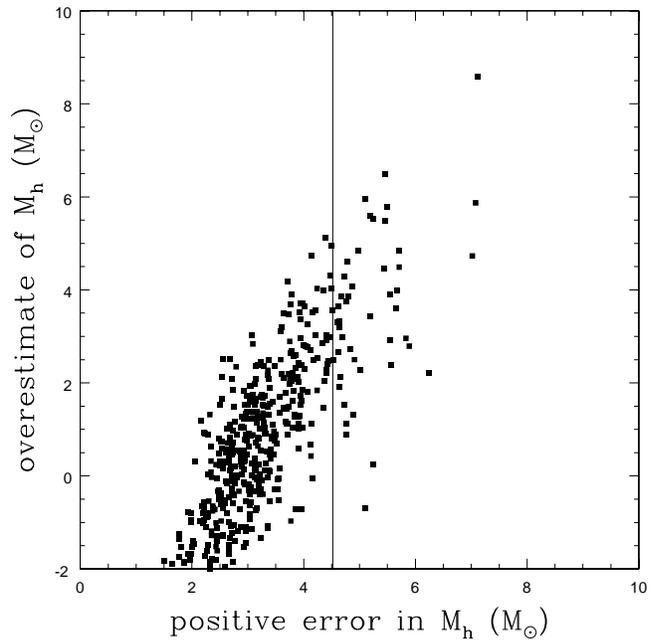}
\caption{
  The correlation between the estimate of the positive
  (upwards) error bar on the Bayesian estimate of $M_h$ and the
  overestimate of $M_h$.  The larger the estimated
  uncertainty, the larger the overestimate of $M_h$.
  A vertical bar shows the positive uncertainty
  estimate for the Bayesian analysis of the actual
  sample.   The \protect\cite{Davies2018} method shows a 
  similar correlation.
  }
\label{fig:over}
\end{figure}

\begin{table}
  \centering
  \caption{Mass Limits for the Progenitor Sample}
  \begin{tabular}{llll}
  \hline
  \multicolumn{1}{c}{Model}    &
  \multicolumn{1}{c}{$M(L)$}   &
  \multicolumn{1}{c}{$M_l/M_\odot$} &
  \multicolumn{1}{c}{$M_h/M_\odot$} \\
\hline
\cite{Smartt2015} &ET04 &$9.5_{-2.0}^{+0.5}$    &$16.5_{-2.5}^{+2.5}$ \\
\cite{Davies2018} &ET04 &$7.5_{-0.2}^{+0.3}$    &$19.0_{-1.3}^{+2.5}$ \\
Davies            &ET04 &$7.49_{-0.27}^{+0.25}$ &$19.05_{-1.30}^{+2.22}$ \\
Bayes             &ET04 &$6.30_{-0.54}^{+0.48}$ &$19.01_{-2.04}^{+4.04}$ \\
\hline 
\cite{Smartt2015} &S18  &$10.0_{-1.5}^{+0.5}$   &$18.5_{-4.0}^{+3.0}$ \\
Davies            &S18  &$8.38_{-0.30}^{+0.28}$ &$21.33_{-1.46}^{+2.48}$ \\
Bayes             &S18  &$7.06_{-0.61}^{+0.54}$ &$21.28_{-2.28}^{+4.52}$ \\
\hline
\multicolumn{4}{l}{The mass-luminosity relation $M(L)$ is either that of ET04} \\
\multicolumn{4}{l}{(\protect\citealt{Eldridge2004}) or S18 (\protect\citealt{Sukhbold2018}) } \\
\multicolumn{4}{l}{where the latter yields a mass about 12\% higher for the } \\
\multicolumn{4}{l}{same luminosity.  The uncertainties are $1\sigma$ except } \\
\multicolumn{4}{l}{for the \protect\cite{Smartt2015} estimates which are at 95\% } \\
\multicolumn{4}{l}{confidence. }  \\
\end{tabular}
\label{tab:results}
\end{table}

\section{Results and Discussion}

To compare these two approaches, we generated 500 simulated
progenitor data sets with minimum and maximum masses of 
$M_l=8M_\odot$ and $M_l= 18M_\odot$ and
then analyzed them using either the approach of \cite{Davies2018}
or the Bayesian method outlined in \S2.  The results
are shown in Figures~\ref{fig:bplot} and \ref{fig:cplot},
respectively.  If we characterize the results by the
median and $1\sigma$ confidence range of the mass estimates, the 
\cite{Davies2018} algorithm finds $M_l/M_\odot= 9.34_{-0.69}^{+0.58}$
and $M_h/M_\odot = 20.01_{-1.84}^{+2.61}$.  The results
are biased to be higher than the input masses, which
is opposite to the sense expected by \cite{Davies2018}.
Our Bayesian approach yields $M_l/M_\odot = 8.44_{-0.65}^{+0.70}$
and $M_h/M_\odot = 18.55_{-1.90}^{+1.95}$, so it is also biased 
to higher masses, but by a smaller amount.  The
scatters in the results for $M_l$ are comparable ($0.61$ 
versus $0.73 M_\odot$, but the Bayesian estimates of 
$M_h$ show significantly less scatter 
($2.62$ versus $2.07M_\odot$). Using different
mass limits produce similar results. For example,
if we raise the upper limit to $M_h= 21 M_\odot$,
we find $M_h/M_\odot = 22.95_{-1.71}^{+2.12}$ for the
\cite{Davies2018} method and $21.27_{-1.87}^{+2.68}$ for 
the Bayesian method.  

The mass estimates from the two statistical approaches are strongly
correlated, as shown in Figure~\ref{fig:compare}.  
Simulated data which leads to an overestimate of $M_h$
by one method also produces an overestimate by
the other method, but the Bayesian method produces
mass estimates systematically closer to the input values.  Examining the
cases with the highest mass estimates, there is a 
fairly general pattern.  The highest mass model star
producing a magnitude measurement has a mass close
to $M_h$.  The randomly selected distance modulus,
extinction, bolometric correction, and magnitude
error combine to produce a model magnitude that is
brighter than the magnitude that would be found using
the nominal values for these quantities.  Then, when
the model is fit to estimate $M_h$, the solutions are
biased high. 

The systems with large mass uncertainties are also the reason why
the additional upwards correction added by \cite{Davies2018}
should not be included.  If the mass uncertainties are 
sufficiently small, then the value of $M_h$ estimated from
a finite sized sample will be an underestimate of the true
limit as they argue.  But this holds only until the typical offset of the 
highest mass progenitor in the sample from the true upper
limit is comparable to the uncertainties in the masses.
Once the uncertainties are larger, the analysis is subject
to a form of Malmquist bias, where it becomes increasingly likely
that an intrinsically lower mass (or equivalently, lower luminosity)
star will be interpreted as
a star above the true upper mass limit.  Based on our
Monte Carlo simulations, this appears to be the regime
appropriate to the existing progenitor data.

For both approaches, there is a correlation between
the uncertainties in $M_h$ and the degree to which $M_h$
is overestimated, as shown in Fig.~\ref{fig:over} for
the Bayesian simulations.  The larger the uncertainties
towards larger masses, the more the mean of the estimator
is biased high. For the $M_h=18M_\odot$ simulations 
and the uncertainties estimated for the actual progenitor
sample below, these correlations would predict that the
\cite{Davies2018} estimate is biased high by 
$(2.6 \pm 1.6)M_\odot$ and the Bayesian estimate
is biased high by $(3.3\pm 0.8)M_\odot$. 

Finally, if we analyze the actual progenitor data, we find
the results given in Table~\ref{tab:results}.  We include
the estimates from \cite{Smartt2015} and \cite{Davies2018}
using the \cite{Eldridge2004} mass-luminosity relation 
(labeled ET04) and then the results using both our 
implementation of the \cite{Davies2018} method (labeled
Davies) and the Bayesian method (labeled Bayes).  We
then repeat the results on the \cite{Sukhbold2018} scale
(labeled S18), although these are simply an offset in 
the mass scale.  Including our knowledge that these
analyses yield estimates of $M_H$ that are biased to
be high, we see that the existence of the red supergiant
problem is quite secure unless the maximum mass of stars
that undergo core collapse as red supergiants can be 
driven below $20 M_\odot$.

\section*{Acknowledgments}

We thank Ben Davies for his comments.
CSK is supported by NSF grants AST-1908570 and AST-1814440.

\end{document}